\documentclass[conference, a4paper]{IEEEtran}
\usepackage[T1]{fontenc}
\IEEEoverridecommandlockouts
\makeatletter
\def\ps@headings{%
\def\@oddhead{\mbox{}\scriptsize\rightmark \hfil \thepage}%
\def\@evenhead{\scriptsize\thepage \hfil \leftmark\mbox{}}%
\def\@oddfoot{}%
\def\@evenfoot{}}
\makeatother
\usepackage{subfigure}
\usepackage{colortbl}
\usepackage{bm}
\usepackage{graphicx}
\usepackage{epstopdf}

\usepackage{graphicx}  
\usepackage{url}       

\usepackage{amsmath}   
\usepackage{cite}
\usepackage{extarrows}
\usepackage{amsfonts,amssymb}

\usepackage{stfloats}

\usepackage{amsfonts}

\usepackage{cases}
\usepackage{algorithm}
\usepackage{multirow}
\usepackage{algorithmic}
\usepackage{geometry}
\newcommand{\ls}[1]
    {\dimen0=\fontdimen6\the\font
     \lineskip=#1\dimen0
     \advance\lineskip.5\fontdimen5\the\font
     \advance\lineskip-\dimen0
     \lineskiplimit=.9\lineskip
     \baselineskip=\lineskip
     \advance\baselineskip\dimen0
     \normallineskip\lineskip
     \normallineskiplimit\lineskiplimit
     \normalbaselineskip\baselineskip
     \ignorespaces
    }

\IEEEoverridecommandlockouts\IEEEpubid{\makebox[\columnwidth]{ 978-1-6654-3540-6/22 \$31.00~\copyright~2022 IEEE \hfill} \hspace{\columnsep}\makebox[\columnwidth]{ }}
\begin{document}

\title{Adaptive Finite Blocklength for Low Access\\ Delay in 6G Wireless Networks}

\author{\IEEEauthorblockN{Yixin Zhang$^{\dagger}$, Wenchi Cheng$^{\dagger}$, and Wei Zhang$^{\ddagger}$}~\\[0.2cm]
\vspace{-10pt}

\IEEEauthorblockA{$^{\dagger}$State Key Laboratory of Integrated
Services Networks, Xidian University, Xi'an, China\\
$^{\ddagger}$School of Electrical Engineering and Telecommunications, The University of New South Wales, Sydney, Australia\\
E-mail: \{\emph{yixinzhang@stu.xidian.edu.cn}, \emph{wccheng@xidian.edu.cn},
\emph{w.zhang@unsw.edu.au}\}}

\vspace{-20pt}

\thanks{This work was supported in part by the National Key Research and Development Program of China under Grant 2021YFC3002102 and in part by the Key R\&D Plan of Shaanxi Province under Grant 2022ZDLGY05-09.
}

}

\maketitle

\begin{abstract}
As the number of real-time applications with ultra-low delay requirements quickly grows, massive ultra-reliable and low-latency communication (mURLLC) has been proposed to provide a wide range of delay-sensitive services for the sixth generation (6G) wireless networks. However, it is difficult to meet the stringent delay demand of massive connectivity with existing grant-based (GB) random access and fixed frame structure in long-term evolution (LTE) and the fifth generation (5G) new radio (NR) systems. To solve this problem, in this paper we propose the new grant-free (GF) based adaptive blocklength scheme for short packet transmission to reduce the access delay. We develop the adaptive blocklength framework where the blocklength can be adaptively changed according to the real-time load, to revise the traditional non-flexible frame structure which impacts the delay performance. Taking the features of mURLLC into consideration, we analyze the GF random access procedure, packet arrival behavior, packet collision, and packet transmission error in the finite blocklength (FB) regime. On this basis, we derive the closed-form expression of successful access and transmission probability and give the GF-based status update model. Then, we propose the access delay minimization problem that jointly considers queuing delay and transmission delay to reduce the overall access delay. With the alternating optimization algorithm, we obtain the optimal blocklength of each packet, thus forming the corresponding adaptive blocklength scheme for mURLLC. Simulation results verify the correctness of theoretical results and show that our proposed adaptive blocklength scheme can significantly reduce the access delay compared with that of LTE and 5G NR systems.

\end{abstract}

\vspace{10pt}

\begin{IEEEkeywords}
mURLLC, grant-free random access, adaptive finite blocklength framework, access delay minimization.
\end{IEEEkeywords}


\section{Introduction}
\IEEEPARstart{W}{ith} the rapid development of Internet of Things (IoT), real-time communications generated by various vertical IoT use cases are emerging in large numbers. Ultra-reliable and low-latency communications (URLLC) has become the core service in the fifth generation (5G) communication networks to meet the ultra-high quality of service (QoS) requirements for various delay-sensitive services. In addition, due to the large number of devices in the IoT scenario, massive access is essential for the IoT system. Integrating URLLC with massive access, massive URLLC (mURLLC) has been proposed in the sixth generation (6G) wireless networks to put forward more stringent requirements on efficient, delay-bounded, and reliable communications~\cite{mURLLC2}. However, in the existing network architecture, especially in uplink transmission, the delay constraints of mURLLC are difficult to satisfy. As a result, delay timeout under massive access condition remains a difficult problem and time-saving solutions are still very important for future IoT networks.

To meet the massive low-delay requirements in mURLLC, the novel grant-free (GF) random access protocol was proposed in 5G new radio (NR) to accommodate massive access within the delay limit. Different from the traditional grant-based (GB) access with four-way handshaking in long-term evolution (LTE), the grant request step has been removed in two-step GF access, where additional delay and overhead can be avoided. Additionally, the long blocklength, which is used for high-capacity-demanded service, is not suitable for the delay-sensitive application whose packet size is small. As a result, the finite blocklength (FB) is proposed as a promising technology to provide a new possibility for short packet transmission to further reduce delay. Relevant work has been carried out to study the GF random access and the FB scheme~\cite{GF1,GF2,FB}. Using the mean-field evolutionary game, the authors investigated the age of information (AoI) minimization problem for GF access~\cite{GF1}. The authors applied deep reinforcement learning (DRL) for GF non-orthogonal multiple access (NOMA) systems, which can mitigate collisions and improve system throughput in an unknown network environment~\cite{GF2}. As for FB, the authors obtained the approximate maximum achievable rate and the packet error rate in the FB regime~\cite{FB}.

Even though various investigations on the benefits of using GF and FB for delay-bounded communications have been carried out, they are almost to separately design GF and FB schemes without combining them. On the one hand, the transmission error probability is no longer close to $0$ in the FB regime, which will impact the GF random access. On the other hand, the potential packet collisions and retransmissions incurred by GF random access also affect the FB analysis. Therefore, the delay performance of the GF and BF combined scheme needs to be reanalyzed. In addition, the existing frame structure design of LTE and 5G NR is based on the fixed transmission time interval (TTI), i.e., fixed blocklength. However, the blocklength has an opposite effect on transmission delay and queuing delay, which means a fixed blocklength will result in an imbalance between the above two delay components, thus leading to an increase in the overall delay~\cite{Xiao}. To fully consider the overall impact of GF and FB on the delay performance, the GF-BF combination analysis and blocklength adjustment scheme with joint delay components optimization are highly demanded.%

To solve this problem, in this paper we propose the new GF-based adaptive blocklength framework for short packet transmission to reduce access delay by combining GF and FB. We first study the GF random access procedure and the corresponding packet arrival behavior. To improve the traditional frame structure, we develop the adaptive blocklength framework where the blocklength can be adaptively changed according to the real-time load. Taking the features of massive users with short packets into consideration, we analyze the packet collision and packet transmission error in the FB regime to derive the closed-form expression of successful access and transmission probability. Then, we model the GF access behavior of each user as a discrete-time Markov chain to establish the GF-based status update system and obtain the corresponding steady-state probability. On this basis, we propose the average access delay minimization problem that jointly considers queuing delay and transmission delay. Then, we propose the alternating optimization (AO) algorithm to solve this non-convex problem. Simulation results show that our developed adaptive blocklength scheme can significantly reduce the access delay for GF random access in the mURLLC scenario.

The rest of this paper is organized as follows. Section~\ref{sec:sys} introduces the GF random access system model. Section~\ref{sec:AFB} proposes the adaptive blocklength framework and the GF-based status update system in the FB regime. Section~\ref{sec:min} presents the access delay minimization problem and the corresponding algorithm. Section~\ref{sec:result} provides the numerical results. Finally, we conclude this paper in Section~\ref{sec:con}.

\section{System Model}\label{sec:sys}

\addtolength{\topmargin}{0.2in}

\begin{figure}[htbp]
\centering\includegraphics[width=3in]{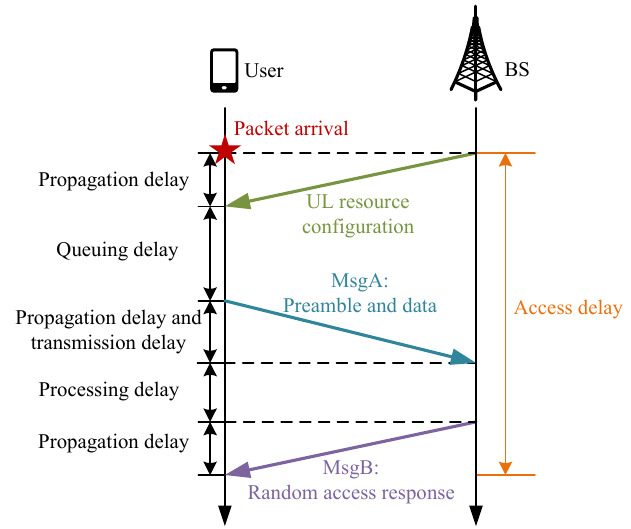}
\caption{The grant-free random access system model.}\label{fig:GF}
\end{figure}
\subsection{Grant-free Random Access Procedure}

Different from GB random access, the active user does not need to wait for the scheduling grant from the base station (BS) in GF random access. That is, once a user becomes active, it directly transmits data with a preamble and waits for the acknowledgment (ACK) from the BS, which can omit the grant step to reduce delay. For contention resolution (CR) and delay limit, the active user starts a CR timer when transmitting a packet. A successful packet transmission of the user is finished if ACK is received before the CR timer expires.

As shown in Fig.\ref{fig:GF}, we consider the access delay as the time interval from each request of the active user is generated until it is successfully transmitted to the BS. The single access delay in GF random access, which means the packet is successfully transmitted at once without any packet collision, denoted by $D^{\rm Acc}_{\rm Single}$, can be expressed as

\begin{equation}
	D^{\rm Acc}_{\rm Single} = D^{\rm Que} + D^{\rm Tra}  + 3D^{\rm Prop} + D^{\rm Proc},
\end{equation}
where $D^{\rm Que}$, $D^{\rm Tra}$, $D^{\rm Prop}$ and $D^{\rm Proc}$ denote the queuing delay, the transmission delay, the propagation delay and the processing delay, respectively. However, GF random access incurs potential collisions because the channels are no longer reserved for a specific packet. When the packet experiences collisions, both Step 1 (Msg A) and Step 2 (Msg B) should be repeated until the packet is successfully transmitted. If collisions occur, the total access delay, denoted by $D^{\rm Acc}$, can be obtained as $D^{\rm Acc} = M^{\rm Re}D^{\rm Acc}_{\rm Single}$, where $M^{\rm Re}$ represents the number of retransmissions until the packet is successfully transmitted within the CR timer.

\subsection{Transmission Model}
We consider an mURLLC system with one BS and $K$ users, where each user experiences both large-scale fading and small-scale fading. The small-scale Rayleigh fading coefficient, denoted by $h$, is with zero mean and unit variance, i.e., $\mathbb E(|h|^2)=1$. In many applications of mURLLC, users are either static or have low mobility \cite{Channel1}, where the duration of each frame is less than the channel coherence time. Thus, the channel is quasi-static and $h$ can be regarded as a constant over several frames. We assume that all users use full path-loss inversion power control with a threshold $P_0$~\cite{P0}. That is, each user controls its transmit power to guarantee the average signal power received at the BS is equal to a predetermined value $P_0$. Considering that it is difficult for users to obtain instantaneous CSI, users use statistical CSI for inversion power control, where statistical CSI is constant over a long period of time and is easy to obtain. In this way, the transmit power of the $k$-th user is $P_k = \frac{d_k^{-\alpha}}{\rho_0}P_0$, where $d_k$, $\alpha$ and $\rho_0$ denote the distance between the $k$-th user and the BS, the pass-loss exponent and the channel power gain at a reference distance of $1$ meter, respectively. Thus, the signal-to-noise ratio (SNR) at the BS can be written as $\gamma = \frac{P_0|h|^2}{\sigma^2}$, where ${\sigma^2}$ denotes the noise power.

\subsection{Pakcet Arrival Model}
We consider a general mURLLC scenario, where each user sporadically generates its short packets. Accordingly, we use the Poisson process as a traffic model~\cite{Possion}, where the user generates new packets following the Poisson distribution with an arrival rate of $\lambda$. Thus, the probability mass function (PMF) of newly arrived packets generated by the $k$-th user can be expressed as
\begin{equation}
	p^{\rm gen}_{k}(a_k)= \Pr\{K = a_k\} = e^{-\lambda T_{\rm max}} \frac{(\lambda T_{\rm max})^{a_k}}{a_k!},
\end{equation}
where $a_k$ represents the number of packets generated by the $k$-th user within a given time interval $T_{\rm max}$.

\section{Adaptive Blocklength Framework}\label{sec:AFB}
For short packet transmission in the mURLLC scenario, the infinite blocklength based on Shannon capacity is no longer applicable. As a result, we perform the FB analysis and propose the adaptive blocklength framework to further increase the system flexibility.
\subsection{Blocklength Structure And The Achievable Rate}
The code length no longer approaches infinity in the FB regime, the blocklength, which refers to the number of symbols transmitted in a frame, denoted by $n$, can be expressed as
\begin{align}\label{eq:TB}
	n=TW,
\end{align}
where $T$ represents the time span (TTI) and $W$ represents the bandwidth resource occupied by the current block. Using the normal approximation \cite{FB2}, the achievable rate in the FB regime, denoted by $R(n)$, can be approximated as 
\begin{align}\label{eq:FBR}
	R(n) \approx {\log _2}\left(1 + \gamma \right) - \sqrt {\frac{V}{n}} Q^{ - 1}(\varepsilon){\log _2}e,
\end{align}
where $V=1-(1+\gamma)^{-2}$ denotes the channel dispersion, $\varepsilon$ denotes the error transmission probability and $Q^{-1}(\varepsilon)$ denotes the inverse of Q-function $Q(\varepsilon)$.

\subsection{Adaptive Blocklength Design}
In the FB regime, there is a new tradeoff relationship of blocklength between transmission delay and queuing delay. 
Based on Eq.~(\ref{eq:FBR}), a short blocklength leads to a small achievable rate, i.e., a small service rate, which results in a large queuing delay. However, a short blocklength corresponds to a small transmission delay. Therefore, the transmission delay and the queuing delay will change in opposite direction with the blocklength changing. For each real-time data load case, there must be an optimal blocklength that can minimize the sum of the transmission delay and the queuing delay according to the different number of users, packet arrival rate and bit number per packet. Therefore, we propose an adaptive blocklength framework to flexibly change the blocklength of each packet based on network load conditions to minimize the total access delay. We assume that the orthogonal bandwidth allocation scheme is employed and the bandwidth is equally allocated to each user. As for TTI, it can be continuously changed with the actual state, thus the adaptive blocklength matrix, denoted by $\boldsymbol n \in \mathbb R ^{K \times Q}$, can be expressed as

\begin{equation}
	\boldsymbol n = W \boldsymbol T =
	\begin{pmatrix}
		n_{1,1}  & \dots & n_{1,q} & \dots &n_{1,Q}\\
		\dots  & \dots & \dots & \dots & \dots\\
		n_{k,1}  & \dots & n_{k,q} & \dots &n_{k,Q}\\
		\dots  & \dots & \dots & \dots & \dots\\
		n_{K,1} & \dots & n_{K,q} & \dots &n_{K,Q}\\
	\end{pmatrix},
\end{equation}
where $Q$ denotes the maximum number of packets sent by the users, $\boldsymbol T \in \mathbb R ^{K \times Q}$ denotes the TTI design matrix and $n_{k,q} = W T_{k,q}$ denotes the blocklength of the $k$-th user with the $q$-th packet.

\subsection{Packet Error Probability}
As the blocklength is finite, a significant packet error probability is introduced, which greatly impairs the access delay performance in the mURLLC system. In other words, even if a collision is avoided, the transmitted packet may still not be successfully received by the BS. According to~\cite{pe_appro_new}, in the FB regime, the packet error probability of the $k$-th user with the $q$-th packet can be tightly approximated to be a linear function as follows:

\begin{equation}
	\varepsilon_{k,q} \approx
	\begin{cases}
		1,& \gamma \leq \gamma_1,\\
		\frac{1}{2}-\mu (\gamma-\beta),& \gamma_1 < \gamma \leq \gamma_2,\\
		0,& \gamma > \gamma_2,\\
	\end{cases}
\end{equation}
where $\mu = \frac{1}{2 \pi }\sqrt{\frac{n_{k,q}}{(2^{2B/n_{k,q}}-1)}}$, $\beta = 2^{B/n_{k,q}} - 1$, $\gamma_1 = \beta - \frac{1}{2\mu}$, $\gamma_2 = \beta + \frac{1}{2\mu}$, and $B$ denotes the bit number of one short packet. Therefore, the packet error probability of the $k$-th user with the $q$-th packet, denoted by $p^{\rm err}_{k,q}$, can be derived as follows:
\begin{equation}
	\begin{split}
		p^{\rm err}_{k,q}&=\mathbb{E}[\varepsilon_{k,q}]\\
		&= \int_{0}^{\gamma_1} f_{\gamma}(x) \mathrm{d}x + \int_{\gamma_1}^{\gamma_2} \Big( \frac{1}{2}-\mu (x-\beta) \Big) f_{\gamma}(x) \mathrm{d}x \\
		&=1-\mu\frac{P_0}{\sigma^2}\Big(e^{-\frac{\sigma^2}{P_0}(\beta - \frac{1}{2\mu})}-e^{-\frac{\sigma^2}{P_0}(\beta + \frac{1}{2\mu})}\Big),\label{Eq:pac-err}
	\end{split}
\end{equation}	
where $f_{\gamma}(x) = \frac{\sigma^2}{P_0}e^{(-\frac{\sigma^2 x}{P_0})}$ for $x \geq 0$ is the probability distribution function (PDF) of the SNR.

\subsection{Packet Collision Avoidance Probability}
In the GF random access, users first randomly select a preamble from the preamble pool containing $M^{\rm Pre}$ preambles. The collision occurs if more than one user selects a same preamble. That is, the user can successfully access the BS by selecting a preamble if only the preamble is selected only by this user. Thus, the packet collision avoidance probability of an active user that the user selects any preamble as long as it is not selected by other users, denoted by $p^{\rm one}$, can be calculated as follows:
\begin{equation}
	p^{\rm one} = M^{\rm Pre} \Big(\frac{1}{M^{\rm Pre}}\Big) \Big(1-\frac{1}{M^{\rm Pre}}\Big)^{K-1}.
\end{equation}

\subsection{Successful Access and Transmission Probability}
The packet can be successfully transmitted only when it selects the unique preamble and the transmission error does not occur. Therefore, the successful access and transmission probability of $k$-th user with the $q$-th packet, denoted by $p^{\rm suc}_{k,q}$, can be derived as
\begin{equation}
	p^{\rm suc}_{k,q} = \Big(1- p^{\rm err}_{k,q}\Big)p^{\rm one} = \Big(1-\mathbb E[\varepsilon_{k,q}] \Big) p^{\rm one}.
\end{equation}

\subsection{Status Update Model}
\begin{figure}[htbp]
	\centering\includegraphics[width=3.3in]{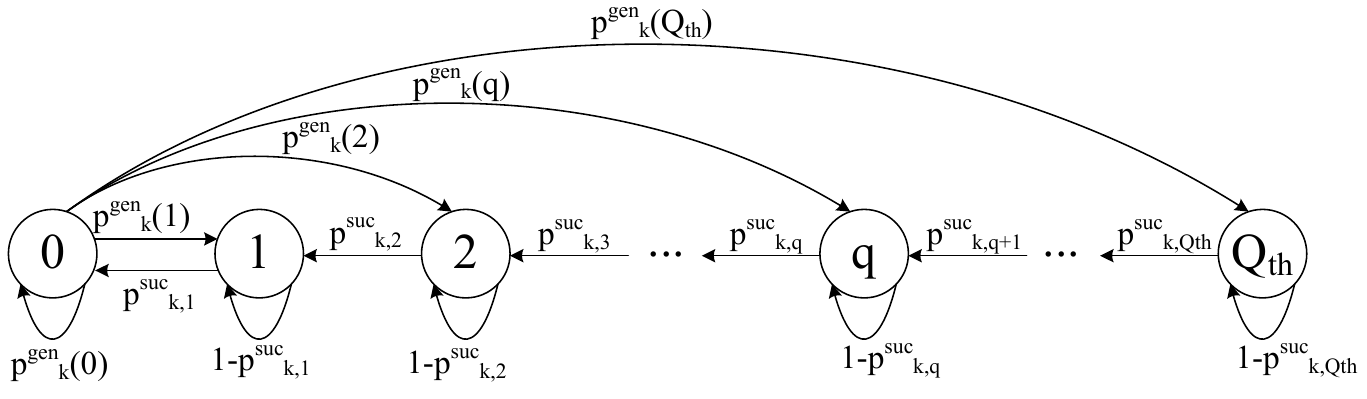}
	\caption{The Markov chain based status update model.}\label{fig:MarkovChain}
\end{figure}
Since the GF random access is considered in this paper, the behavior of each user can be modeled as a discrete-time Markov chain process as shown in Fig.~\ref{fig:MarkovChain}. Each state represents the queuing length of each user, where the queuing length implies the number of packets in the queue. The state space $\mathcal{S}$ can be expressed as $\mathcal{S}=\{0,1,\cdots q ,\cdots,Q_{\rm th}\}$, where $Q_{\rm th}$ represents the maximum queuing length size. The distribution of the state probability at time $t$, $\boldsymbol\pi(t)$, can be expressed as $\boldsymbol\pi(t)=\{\pi_0(t),\pi_1(t),\cdots, \pi_q(t), \cdots, \pi_{Q_{\rm th}}(t)\}$, where $\pi_q(t) (0\leq q \leq Q_{\rm th})$ represents the state probability that the queuing length equals to $q$ at time $t$. A state transition occurs whenever each user attempts to perform uplink transmission. 
In order to mathematically derive the access delay, we should obtain the stationary distribution $\boldsymbol\pi$. According to the state transition shown in Fig.~\ref{fig:MarkovChain}, the steady-state probability of the $k$-th user can be expressed as

\begin{equation}\label{Eq:state}
	\left\{
	\begin{aligned}
		&\pi_{k,0} = \frac{ \prod_{i=1}^{Q_{\rm th}} p^{\rm suc}_{k,i}}{ \prod_{i=1}^{Q_{\rm th}} p^{\rm suc}_{k,i} + \sum_{j=1}^{Q_{\rm th}} \Big[\prod_{r=1,r \neq j}^{Q_{\rm th}} p^{\rm suc}_{k,r} \sum_{l=j}^{Q_{\rm th}} p^{\rm gen}_k(l) \Big] }, & & \\
		& \pi_{k,q} = \frac{\pi_{k,0} \Big(\sum_{l=q}^{Q_{\rm th}}p^{\rm gen}_k(l)\Big)}{p^{\rm suc}_{k,q}},  1 \leq q \leq Q_{\rm th}.&&
	\end{aligned}
	\right.
\end{equation}
It can be seen from Eq.~(\ref{Eq:state}) that the steady-state probability distribution is a function of the successful access and transmission probability of each user at each attempt.

\section{Access Delay Minimization Problem}\label{sec:min}
To satisfy mURLLC stringent delay demands, we propose the access delay minimization problem that considers both queuing delay and transmission delay to reduce the overall access delay by adjusting the blocklength of each packet.
\subsection{Access Delay Reformulation}
Based on the above adaptive blocklength framework, we reformulate the expression of the access delay. The queuing delay refers to the time duration between the new packet arrival and the first attempt of GF uplink transmission, thus the queuing delay of the $k$-th user, denoted by $D^{\rm Que}_k$, can be expressed as
\begin{equation}
	D^{\rm Que}_k = \sum_{q=1}^{Q_{\rm th}} \pi_{k,q} \cdot D^{\rm Que}_{k,q}, 1 \leq k \leq K,
\end{equation}
where $D^{\rm Que}_{k,q}$ represents the queuing delay of a new packet arrives when the queuing length is equal to $q$, which can be written as
\begin{equation}
	D^{\rm Que}_{k,q} = \sum_{l=0}^{q-1} \Big(T_{k,l}  + D^{\rm P}\Big)\cdot \mathbb{E}[X_{k,l}], 1 \leq k \leq K, 1 \leq q \leq Q_{\rm th},
\end{equation}
where $D^{\rm P}=3D^{\rm Prop} + D^{\rm Proc}$ and $X_{k,q}$ represents the number of retransmissions until the $q$-th packet of the $k$-th user is successfully transmitted, which follows a geometric distribution with parameter $p^{\rm suc}_{k,q}$~\cite{Stochastic} and its PMF can be expressed as
\begin{equation}
	f_{X_{k,q}}(x)=p^{\rm suc}_{k,q}(1-p^{\rm suc}_{k,q})^{x-1}, x\in\{1,2,3,\cdots\}.
\end{equation}
The transmission delay refers to the time duration from the first attempt of GF uplink transmission to the ACK message is successfully received. The transmission delay of the $k$-th user, denoted by $D^{\rm Tra}_k$, can be written as
\begin{equation}
	D^{\rm Tra}_k =  \sum_{q=0}^{Q_{\rm th}} \Big(T_{k,q} + D^{\rm P} \Big) \cdot \mathbb{E}[X_{k,q}], 1 \leq k \leq K.
\end{equation}
Therefore, the access delay of the $k$-th user, denoted by $D^{\rm Acc}_k$, can be expressed as
\begin{equation}
	\begin{aligned}
	D^{\rm Acc}_k = &\sum_{q=1}^{Q_{\rm th}} \pi_{k,q} \sum_{l=0}^{q-1} \Big(T_{k,l} + D^{\rm P}\Big)  \cdot \mathbb{E}[X_{k,l}] + \\
	& \ \ \  \sum_{q=0}^{Q_{\rm th}} \Big(T_{k,q} + D^{\rm P}\Big) \cdot \mathbb{E}[X_{k,q}], 1 \leq k \leq K.  \\
	\end{aligned}
\end{equation}

\subsection{Access Delay Minimization Problem Formulation}
The average access delay minimization problem, denoted by $\textbf{\textit{P}1}$, can be expressed as
\begin{align}\label{P1}
	\textbf{\textit{P}1:\ } &\min_{  \boldsymbol n}\ \ D^{\rm Acc}_{\rm ave}= \frac{1}{K}\sum_{k=1}^{K} D^{\rm Acc}_k \tag{16a} \\ 
	\  \mathrm{s.t.}\ \  &1).\ n_{k,q} \geq 0, 1 \leq k \leq K, 0 \leq q \leq Q_{\rm th},\tag{16b}\\ 
		\ \  &2).\ \sum_{q=0}^{Q_{\rm th}} R(n_{k,q}) n_{k,q} \geq B \sum_{a_k=0}^{Q_{\rm th}} p^{\rm gen}_{k}(a_k)a_k , 1 \leq k \leq K,\tag{16c}\label{Eq:constranit3}
\end{align}
where $D^{\rm Acc}_{\rm ave}$ denotes the average access delay of $K$ users. 
However, both the objective function and the constraint Eq.~(\ref{Eq:constranit3}) are non-convex. 
To solve this problem, we employ the AO algorithm. That is, we optimize the $q$-th blocklength under the condition of fixing $h$-th blocklength, where $h\in \mathcal{H}, \mathcal{H} = \{0 \leq h \leq Q_{\rm th}, h\neq q\}$. Thus, the variable to be optimized becomes ${\boldsymbol n_q (0 \leq q \leq Q_{\rm th}})$.

\subsection{Problem Transformation}
To convert the constraints to be convex, we adopt the penalty function method to move Eq.~(\ref{Eq:constranit3}) into the objective function, where $\textbf{\textit{P}1}$ can be transformed into $\textbf{\textit{P}2}$ as follows:
\begin{align}
	\textbf{\textit{P}2:\ } & \min_{\boldsymbol n_q }\ \ \frac{1}{K}\sum_{k=1}^{K} \Bigg (\sum_{q=1}^{Q_{\rm th}} \pi_{k,q} \sum_{l=0}^{q-1} \Big(T_{k,l} + D^{\rm P}\Big) \mathbb{E}[X_{k,l}] +\nonumber\\  
	& \hspace{0.6cm}  \sum_{q=0}^{Q_{\rm th}} \Big(T_{k,q} + D^{\rm P}\Big)\mathbb{E}[X_{k,q}] \Bigg) + \omega \Big[\min \big\{g(n_{k,q}),0\big\}\Big]^2 \tag{17a}\\ 
	\  &\mathrm{s.t.}\ \  \ n_{k,q} \geq 0, 1 \leq k \leq K, 0 \leq q \leq Q_{\rm th},\tag{17b}
\end{align}
where $\omega$ is the penalty factor, $g(n_{k,q}) = \sum_{q=0}^{Q_{\rm th}} R(n_{k,q}) n_{k,q} - B\sum_{a_k=0}^{Q_{\rm th}} p^{\rm gen}_{k}(a_k)a_k$. To deal with the non-convex objective function, we divide it into convex and non-convex parts and replace the variables of them with $\boldsymbol x_q$ and $\boldsymbol y_q$, respectively. Then, $\textbf{\textit{P}2}$ can be converted into $\textbf{\textit{P}3}$ as follows:
\begin{align}
	&\textbf{\textit{P}3:\ } \min_{\left(\boldsymbol x_q, \boldsymbol y_q \right) }\ \ u(\boldsymbol x_q)+v(\boldsymbol y_q) \tag{18a} \\  
	 &\ \mathrm{s.t.}\ \  \boldsymbol x_q - \boldsymbol y_q = \boldsymbol 0, \boldsymbol x_q \in \mathcal{U} \tag{18b},
\end{align}
where $\mathcal{U} = \{\boldsymbol n_q | n_{k,q} \geq 0, 1 \leq k \leq K, 0 \leq q \leq Q_{\rm th}\}$. Then, the corresponding augmented Lagrangian function can be expressed as
\begin{equation}
	\mathcal{L}_q (\boldsymbol x_q, \boldsymbol y_q, \boldsymbol \lambda_q) = u(\boldsymbol x_q) + v(\boldsymbol y_q) + \boldsymbol \lambda^T_q ( \boldsymbol x_q - \boldsymbol y_q ) + \frac{\tau_q}{2}||\boldsymbol x_q - \boldsymbol y_q||^2, \tag{19}
\end{equation}
where $\boldsymbol \lambda_q$ and $\tau_q$ represent the Lagrangian multipliers and the step size parameter. 
Then, we can update $\boldsymbol x_q, \boldsymbol y_q, \boldsymbol \lambda_q$ as follows:
\begin{equation}
	\left\{
	\begin{aligned}
		& \boldsymbol x^{k+1}_q =  \underset{\boldsymbol x_q} {\arg\min} \ \mathcal{L}_q (\boldsymbol x_q, \boldsymbol y^k_q, \boldsymbol \lambda^k_q ; \tau_q), \boldsymbol x_q \in \mathcal{U}, & & \\
		& \boldsymbol y^{k+1}_q =  \underset{\boldsymbol y_q} {\arg\min} \ \mathcal{L}_q (\boldsymbol x^{k+1}_q, \boldsymbol y_q, \boldsymbol \lambda^k_q ; \tau_q), & & \\
		& \boldsymbol \lambda^{k+1}_q = \boldsymbol \lambda^{k}_q + \tau_q (\boldsymbol x^{k+1}_q-\boldsymbol y^{k+1}_q), & & 
	\end{aligned}\tag{20}
	\right.
\end{equation}
where the label in the upper right corner represents the iteration index.
\subsection{The Alternating Optimization Algorithm}
Then, we use the AO algorithm to solve the above problem, where the detailed flow is outlined in Algorithm 1. In the first iteration, we first optimize the blocklengths of the first packet from all users with other blocklengths fixed. Next, we optimize the blocklengths of the following packets until the blocklengths of the ${Q_{\rm th}}$-th packet from all users are optimized. The above output is saved as the first iteration result and the following iterations are performed until the average access delay converges. Finally, the optimal blocklength $\boldsymbol n^*$ and the corresponding minimum average access delay can be obtained, thus forming the adaptive blocklength scheme. 
\begin{algorithm}[h]
	\caption{Alternating optimization for solving $\textbf{\textit{P}1}$.}
	\begin{algorithmic}[1]
		
		\STATE Initialize the AO iteration index $t=0$.
		\REPEAT
		\STATE Initialize the blocklength iteration index $q=0$ and blocklength matrix $\boldsymbol n_q= \boldsymbol n^{(0)}_q$.
		\STATE \textbf{repeat}
		\STATE \hspace{0.4cm} For the given blocklength $\boldsymbol n_h^{(t)}$, optimize $\boldsymbol n_q^{(t+1)}$ \\ \hspace{0.4cm} by solving problem $\textbf{\textit{P}3}$:
		\STATE \hspace{0.4cm} Initialize the inner iteration index $l=0$,\\\hspace{0.4cm}  $\boldsymbol x_q = \boldsymbol x^{(0)}_q$, $\boldsymbol y_q = \boldsymbol y_q^{(0)}$, $\boldsymbol \lambda_q = \boldsymbol \lambda^{(0)}_q$.
		\STATE \hspace{0.4cm} \textbf{repeat}
		\STATE \hspace{0.4cm} \hspace{0.4cm} Update $l\leftarrow l+1$.
		\STATE \hspace{0.4cm} \hspace{0.4cm} Update $\boldsymbol x^{(l)}_q$, $\boldsymbol y^{(l)}_q$, $\boldsymbol \lambda^{(l)}_q$ in turn.
		\STATE \hspace{0.4cm} \textbf{until} $\mathcal{L}_q$ converges.
		\STATE \hspace{0.4cm} Update $q\leftarrow q+1$.
		\STATE \textbf{until} $q=Q_{\rm th}$.
		\STATE Update $t\leftarrow t+1$.
		\UNTIL{$D^{\rm Acc}_{\rm ave}$ converges.}
	\end{algorithmic}
\end{algorithm}

\section{Numerical Results}\label{sec:result}
In this section, we evaluate the performance of our proposed adaptive blocklength scheme for mURLLC low-delay communications. Consider a single-cell cellular network with one BS and $K$ randomly distributed users. We set the bandwidth $W=1$~MHz, the inversion power control threshold $P_{0}=-90$~dBm, the noise variance $\sigma^2=-90$~dBm, the processing and propagation delay $D^{\rm P}=1$~ms and the maximum queuing length $Q_{\rm th}=5$, the number of preambles $M^{\rm Pre} = 20$, respectively. The TTI adopted by LTE and 5G NR in the benchmark scheme is set to $1$~ms and $0.5$~ms.

\begin{figure}[htbp]
	\begin{center}
		\subfigure[The successful access and transmission probability versus the blocklength under the different number of users.]{
			\includegraphics[width=1.5in]{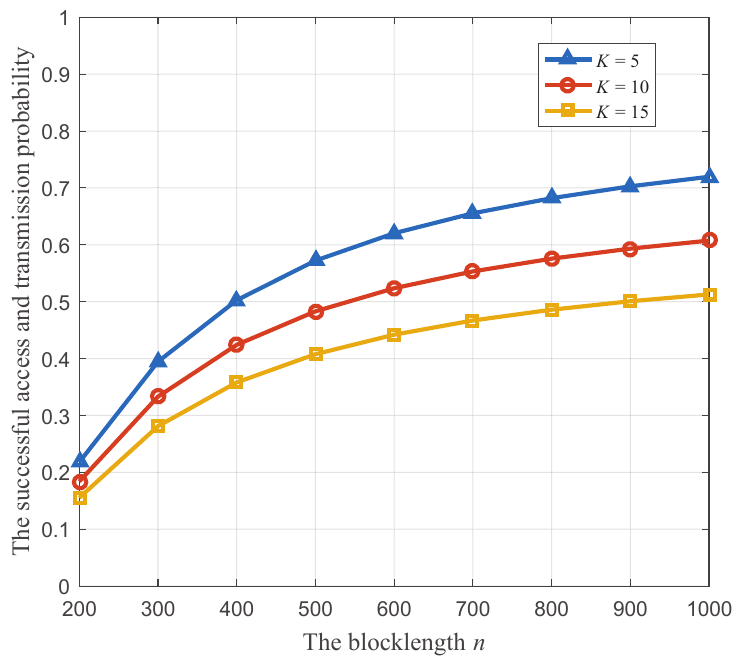}\label{fig:Suc}
		}
		\subfigure[The blocklength of each packet with the adaptive blocklength scheme and the fixed blocklength scheme in LTE and 5G NR.]{
			\includegraphics[width=1.5in]{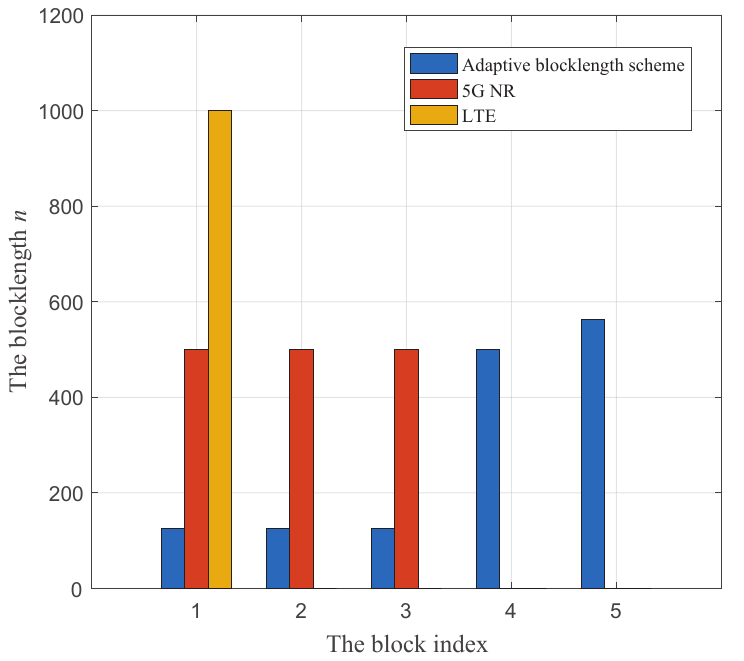}\label{fig:Block}
		}
		\caption{Changes in the successful access and transmission probability and the blocklength of the adaptive blocklength scheme.}
	\end{center}
\end{figure}

Figure~\ref{fig:Suc} plots the successful access and transmission probability versus the blocklength $n$ under the different number of users $K$. The successful access and transmission probability gradually increases as the blocklength $n$ increases. Furthermore, the successful probability decreases as the number of users $K$ increases. This is caused by the mismatch between the number of users and preambles, which means that an appropriate number of preambles should be selected according to the number of users. Fig.~\ref{fig:Block} shows the blocklength of our proposed adaptive blocklength scheme and fixed blocklength scheme in LTE and 5G NR. It can be seen from Fig.~\ref{fig:Block} that the blocklength of LTE and 5G NR remains constant during the whole transmission process, whereas the blocklength of our proposed adaptive blocklength scheme sequentially changes as the block index increases.

\begin{figure}[htbp]
	\centering\includegraphics[width=2.5in]{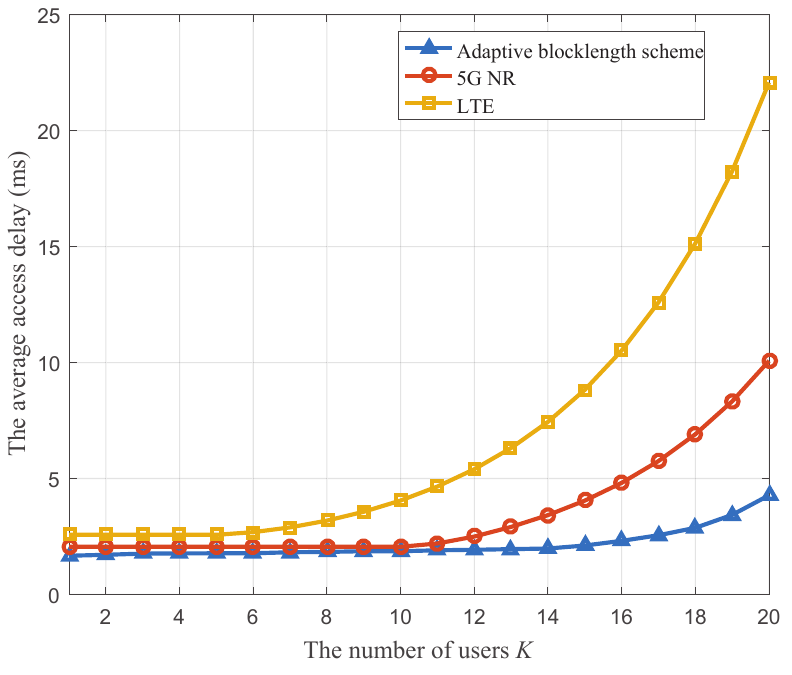}
	\caption{The access delay of the proposed adaptive blocklength scheme versus the number of users compared with that of LTE and 5G NR.}\label{fig:F2}
\end{figure}

Figure~\ref{fig:F2} depicts the average access delay under different numbers of users $K$. The average access delay increases as $K$ increases. This is because as the number of users increases, the probability of multi-user collision increases, where the number of retransmissions increases and the corresponding access delay accordingly increases. However, for a fixed value of $K$, our proposed adaptive blocklength scheme can achieve the lowest average access delay. Besides, with the number of users increasing, the growth rate of the access delay with our proposed scheme is the slowest compared with that of LTE and 5G NR. This proves that our proposed scheme can effectively reduce the access delay and the blocklength can be adaptively adjusted in time to adapt to the real-time load when the number of users $K$ changes. 

\begin{figure}[htbp]
	\centering\includegraphics[width=2.9in]{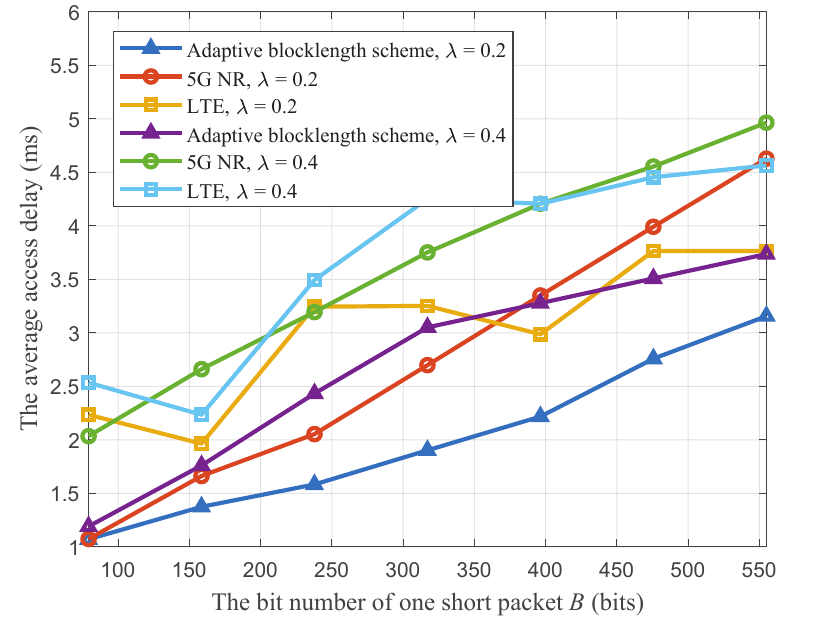}
	\caption{The access delay of the proposed adaptive blocklength scheme versus the bit number of one short packet and arrival rate compared with that of LTE and 5G NR.}\label{fig:F3}
\end{figure}

Figure~\ref{fig:F3} plots the average access delay versus the bit number of one short packet $B$ with two different packet arrival rates $\lambda_1=0.2$ and $\lambda_2=0.4$. The access delay increases with the bit number of one short packet and the arrival rate increasing. In addition, for fixed $B$ and $\lambda$, using our proposed adaptive blocklength scheme, the lowest access delay can be achieved compared to LTE and 5G NR with fixed blocklength.

\section{Conclusion}\label{sec:con}
In this paper, we solved the problem of how to reduce the access delay for mURLLC in 6G wireless networks. Combining GF and FB, we proposed the new GF-based adaptive blocklength framework, where the blocklength can be flexibly changed according to the real-time data load. Considering the massive short packet transmission characteristics of mURLLC, we derived the closed-form expression of successful access and transmission probability and gave the status update model. On this basis, we proposed the access delay minimization problem which fully considers each component of the access delay to reduce the overall access delay. Compared with the fixed blocklength structure in LTE and 5G NR systems, our proposed GF-based adaptive blocklength scheme can significantly reduce the access delay. Meanwhile, our work provided a reference for studying the status update process in the mURLLC system.
\bibliographystyle{IEEEtran}
\bibliography{References-GC}
\end{document}